\documentclass[a4paper]{report}
\usepackage{RJournal}
\usepackage[round]{natbib}
\begin{document}

\fancyhf{}
\fancyhead[LO,RE]{\textsc{Contributed Article}}
\fancyhead[RO,LE]{\thepage}
\fancyfoot[L]{The R Journal Vol. X/Y, Month, Year}
\fancyfoot[R]{ISSN 2073-4859}

\begin{article}

\title{\pkg{StellaR}: a Package to Manage Stellar Evolution Tracks and
Isochrones} 
\author{by Matteo Dell'Omodarme, and Giada Valle}

\maketitle

\abstract{
We present the R package \pkg{stellaR}, which is designed to access and
manipulate publicly available  stellar evolutionary tracks and isochrones
from the Pisa low-mass database.  
The procedures of the extraction of important stages in the evolution of a star
from the database, of 
the isochrones construction from stellar tracks and of the
interpolation among tracks are discussed and demonstrated.
}

Due to the advance in the instrumentation, 
nowadays astronomers can deal with a huge amount of
high-quality 
observational data. In the last decade
impressive  improvements of spectroscopic and photometric
observational capabilities made available data which stimulated the research
in the globular clusters field.
The theoretical effort of recovering the evolutionary history of
the cluster benefits by the computation of extensive databases of stellar tracks
and isochrones, such as 
\citet{teramo06, dotter08, padova08}. We recently computed a large data set 
of stellar tracks and isochrones, ``The Pisa low-mass
database'' \citep{database2012}, with up to date  
physical and chemical inputs, and made available all
the calculations to the 
astrophysical community at the Centre de Donn\'ees astronomiques de Strasbourg
(CDS)\footnote{via anonymous ftp to 
cdsarc.u-strasbg.fr, or via 
\url{http://cdsarc.u-strasbg.fr/viz-bin/qcat?J/A+A/540/A26}},
a data center dedicated to the collection and worldwide distribution of
astronomical data. 

In most databases, the management of the information and the extraction
of the relevant evolutionary properties  from libraries of tracks and/or 
isochrones is the responsibility of the end users. Due to its extensive
capabilities of data manipulation 
and analysis, however, R is an ideal choice for these tasks. Nevertheless R is
not yet well known in astrophysics; up to December 2012 only seven
astronomical or astrophysical-oriented packages have been published  on CRAN
(see the taskview 
Chemometrics and Computational Physics). 

The package \CRANpkg{stellaR} \citep{stellar} is an effort to make
available to the astrophysical community a basic tool-set with the following
capabilities: retrieve the required
calculations from CDS;
plot the information in a suitable form; construct by 
interpolation tracks or isochrones of compositions different to the ones
available in the database; construct isochrones for age not included in the
database; extract relevant evolutionary points from tracks or isochrones.

\section{Get stellar evolutionary data}

The Pisa low-mass database contains computations classified according to four
parameters: 
the metallicity $z$ of the star, its initial helium value $y$, the value of
$\alpha$-enhancement of the heavy elements mixture with respect to the
reference mixture and  
the mixing length parameter $\alpha_{\rm ml}$ used to model external
convection efficiency.
The values of the parameters available in the database can be displayed using
the function \code{showComposition()}:
\begin{example}
> showComposition()
Mixing-length values:
	 1.7, 1.8, 1.9 

alpha-enhancement values:
         0, 1  (i.e. [alpha/Fe] = 0.0 
                     [alpha/Fe] = 0.3) 

Chemical compositions:
      z    y.1   y.2   y.3   y.4   y.5   y.6
  1e-04  0.249  0.25  0.27  0.33  0.38  0.42
  2e-04  0.249  0.25  0.27  0.33  0.38  0.42
  3e-04  0.249  0.25  0.27  0.33  0.38  0.42
  4e-04  0.249  0.25  0.27  0.33  0.38  0.42
  5e-04  0.250  0.25  0.27  0.33  0.38  0.42
  6e-04  0.250  0.25  0.27  0.33  0.38  0.42
  7e-04  0.250  0.25  0.27  0.33  0.38  0.42
  8e-04  0.250  0.25  0.27  0.33  0.38  0.42
  9e-04  0.250  0.25  0.27  0.33  0.38  0.42
  1e-03  0.250  0.25  0.27  0.33  0.38  0.42
  2e-03  0.252  0.25  0.27  0.33  0.38  0.42
  3e-03  0.254  0.25  0.27  0.33  0.38  0.42
  4e-03  0.256  0.25  0.27  0.33  0.38  0.42
  5e-03  0.258  0.25  0.27  0.33  0.38  0.42
  6e-03  0.260  0.25  0.27  0.33  0.38  0.42
  7e-03  0.262  0.25  0.27  0.33  0.38  0.42
  8e-03  0.264  0.25  0.27  0.33  0.38  0.42
  9e-03  0.266  0.25  0.27  0.33  0.38  0.42
  1e-02  0.268  0.25  0.27  0.33  0.38  0.42 
\end{example}
The table of chemical composition presents all the $y$ values available for a
given $z$. 
For a set of parameters, the tracks files are identified specifying 
the mass of the desired model (in the range [0.30, 1.10] $M_\odot$ ($M_\odot
= 1.99 \cdot 10^{33}$ g is the mass of the Sun), step of
0.05 $M_\odot$), while the age (in the range [8.0 - 15.0] Gyr, step of 0.5
Gyr) is required for the isochrones.  

Upon specification of the aforementioned parameters, the \pkg{stellaR}
package 
can import data from CDS (via anonymous ftp)  
from an
active Internet connection. 
The CDS data are stored in ASCII format, and include a header with calculation
metadata, such as the metallicity, the inital helium abundance, the
mixing-length.  
The import is done via a \code{read.table} call, skipping the header of the
files.  

The following data objects can be downloaded from
the database site:
\begin{itemize}
\item{Stellar track: a stellar evolutionary track computed starting from
Pre-Main 
Sequence (PMS) and ending at the onset of helium flash (for masses $M \geq
0.55$ $M_\odot$) or at the exhaustion of central hydrogen (for $0.30 \;
M_\odot \leq M \leq 0.50 \; M_\odot$). The
functions \code{getTrk()} and \code{getTrkSet()} can be used to access such
data; they respectively return objects of classes \code{"trk"}
and \code{"trkset"}. } 
\item{Stellar ZAHB: Zero-Age Horizontal-Branch models. The
functions \code{getZahb()} can be used to access such
data; it returns an object of class \code{"zahb"}.}
\item{HB models: computed from ZAHB to the onset of thermal pulses. The
functions \code{getHb()} and \code{getHbgrid()} can be used to access such
data; they respectively return objects of classes \code{"hb"}
and \code{"hbgrid"}. 
} 
\item{Stellar isochrones: computed in the age range [8 - 15] Gyr. The
functions \code{getIso()} and \code{getIsoSet()} can be used to access such
data; they respectively return objects of classes \code{"iso"}
and \code{"isoset"}.  }
\end{itemize}
Readers interested in details about the computation procedure are referred
to \citet{database2012}. 
The data gathered from CDS are organized into objects of appropriate
classes. The package includes
\code{print} and \code{plot} S3 methods for the classes  \code{"trk"},
\code{"trkset"}, \code{"hb"}, \code{"zahb"}, \code{"hbgrid"},  
\code{"iso"},  
and \code{"isoset"}.   

As an example, we illustrate the recovering of the stellar track for a model
of mass 
$M=0.80$ $M_\odot$, metallicity $z =0.001$, initial helium abundance
$y=0.25$, mixing-length $\alpha_{\rm ml} = 1.90$, $\alpha$-enhancement
[$\alpha$/Fe] = 0.0.
\begin{example} 
> track <- getTrk(m = 0.80, z = 0.001, 
           y = 0.25, ml = 1.90, afe = 0)
> track
	 Stellar track

Mass = 0.8 Msun
Z = 0.001 , Y = 0.25 
Mixing length = 1.9 
[alpha/Fe] = 0 

> names(track)
[1] "mass"      "z"         "y"         "ml"        
    "alpha.enh" "data"

> class(track)
[1] "trk"     "stellar"
\end{example}
The function \code{getTrk()} returns an object of class \code{"trk"},
which is a list containing the track metadata, i.e.\
as the star mass, the metallicity, the initial He abundance, the mixing-length
and the  $\alpha$-enhancement, and the computed data in the data
frame \code{data}.
Track data contains the values of 15 variables:
\begin{example}  
> names(track\$data)
 [1] "mod"     "time"    "logL"    "logTe"   
     "mass"    "Hc"      "logTc"   "logRHOc" 
     "MHEc"    "Lpp"     "LCNO"    "L3a"     
     "Lg"      "radius"  "logg" 
\end{example}
The included variables are: \code{mod} the progressive model
number; \code{time} the logarithm of 
the stellar age (in yr); \code{logL} the logarithm of the surface luminosity
(in unit 
of solar luminosity); \code{logTe} the logarithm of the effective temperature
(in K); \code{mass} the stellar mass (in unit of solar mass); \code{Hc} the
central hydrogen abundance (after hydrogen exhaustion: central helium
abundance); \code{logTc}  the logarithm of the central temperature
(in K); \code{logRHOc}  the logarithm of the central density
(in g/cm$^3$); \code{MHEc}  the mass of the helium core (in units of solar
mass); \code{Lpp} the luminosity of pp chain (in units of surface luminosity);
\code{LCNO} the luminosity of CNO chain (in units of surface luminosity);
\code{L3a} the luminosity of triple-$\alpha$ burning (in units of surface
luminosity); \code{Lg} luminosity of the gravitational energy (in units of
surface luminosity); \code{radius} the stellar radius (in units of solar
radius); \code{logg} the logarithm of surface gravity (in cm/s$^2$).

Similarly the part of the track starting from ZAHB and ending at the onset of
thermal pulses can be downloaded by the call:
\begin{example} 
> hbtk <- getHb(m = 0.80, z = 0.001, 
            y = 0.25, ml = 1.90, afe = 0)
> hbtk
	 Stellar track from ZAHB

Mass = 0.8 Msun
Mass RGB = 0.8 Msun
Z = 0.001 , Y = 0.25 
Mixing length = 1.9 
[alpha/Fe] = 0 

> names(hbtk)
[1] "mass"      "massRGB"   "z"         "y"         
    "ml"        "alpha.enh" "data"

> class(hbtk)
[1] "hb"      "stellar"
\end{example}
which returns an object of class \code{"hb"}, which differs from an object of
class \code{"trk"} only
for the presence of the variable \code{massRGB}, i.e.\ the Red-Giant Branch
(RGB) progenitor mass. 

Usually a set of tracks with different mass and/or metallicity are needed for
computations. The package \pkg{stellaR} provides the
function \code{getTrkSet()}, which can download a set of tracks with different
values of mass, metallicity, helium and mixing-length.  
As an example the whole set of masses (from 0.30 to 1.10 $M_\odot$, step of
0.05 $M_\odot$), for metallicity $z =0.001$, initial helium
abundance 
$y=0.25$, mixing-length $\alpha_{\rm ml} = 1.90$, $\alpha$-enhancement
[$\alpha$/Fe] = 0.0 can be downloaded as follows:  
\begin{example} 
> mass <- seq(0.3, 1.1, by = 0.05)
> trks <- getTrkSet(m = mass, z = 0.001, 
             y = 0.25, ml = 1.90, afe = 0)
> trks
[[1]]
	 Stellar track

Mass = 0.3 Msun
Z = 0.001 , Y = 0.25 
Mixing length = 1.9 
[alpha/Fe] = 0

[[2]]
	 Stellar track

Mass = 0.35 Msun
Z = 0.001 , Y = 0.25 
Mixing length = 1.9 
[alpha/Fe] = 0

...
\end{example} 

\begin{figure}
\includegraphics[width=8cm]{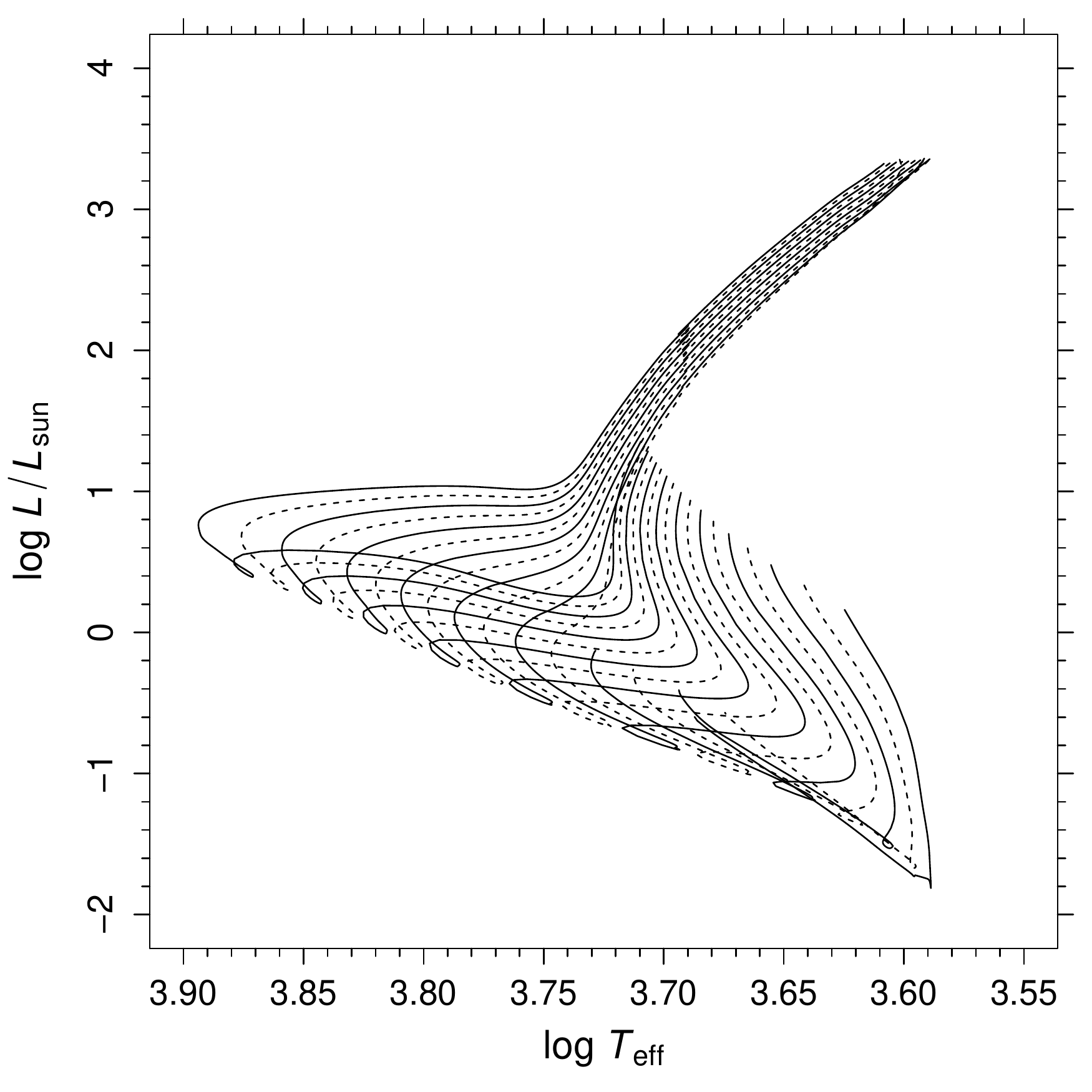}
\caption{\label{fig:trkset}
The evolutionary tracks for masses from $M = 0.30$ $M_\odot$ to  $M = 1.10$
$M_\odot$ from PMS to He flash. The parameters of the calculations are: $z$ =
0.001, $y$ = 0.25, $\alpha_{\rm ml}$ = 1.90, [$\alpha$/Fe] = 0.0.}
\end{figure}

The function \code{getTrkSet()} returns an object of class \code{"trkset"}, a
list containing objects of 
class \code{"trk"}. The track set can be displayed in the usual ($\log T_{\rm
eff}$, $\log L/L_\odot$) plane by a call of the function \code{plot()}:
\begin{example} 
> plot(trks, lty = 1:2)
\end{example} 
The output of the function is shown in Figure \ref{fig:trkset}. The plot is
produced by a call to the function \code{plotAstro()} which allows the user to 
customize several aspects of the plot, such as the axes label, the number of
minor ticks between two major ticks, the limits of the axes, the color and
type of the lines (as in the example), the type of the plot (lines, points,
both, \ldots).

\begin{figure}
\includegraphics[width=8cm]{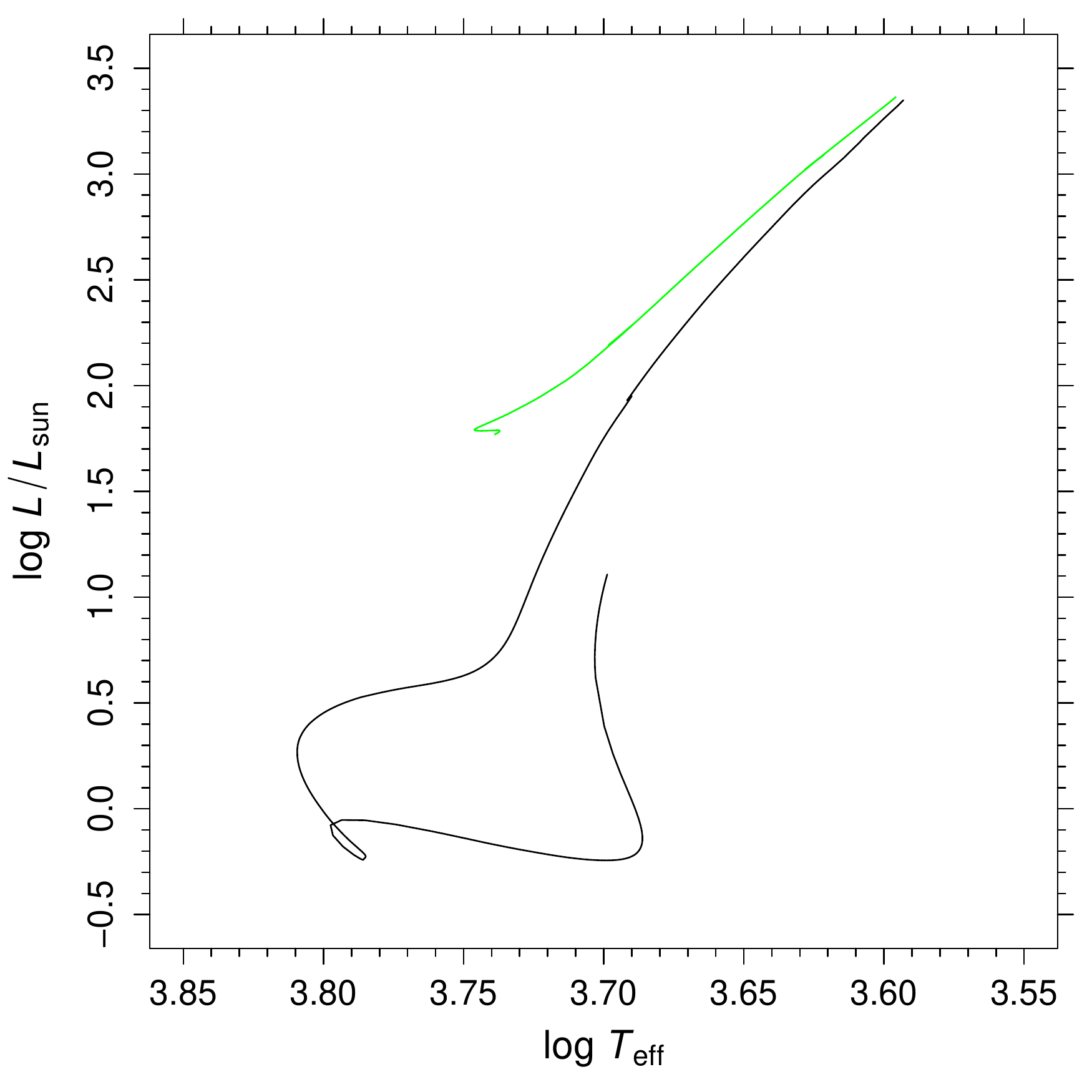}
\caption{\label{fig:trkhb}
(black line): evolutionary tracks for mass $M = 0.80$ $M_\odot$, $z$ =
0.001, $y$ = 0.25, $\alpha_{\rm ml}$ = 1.90, [$\alpha$/Fe] = 0.0.
from PMS to He flash. (green line): evolutionary track from ZAHB to thermal
pulses. }
\end{figure}

The use of \code{plot()} function is further demonstrated in
Figure \ref{fig:trkhb}, where, for $z$ =
0.001, $y$ = 0.25, $\alpha_{\rm ml}$ = 1.90, [$\alpha$/Fe] = 0.0, are
displayed the evolutionary track for $M = 
0.80$ $M_\odot$ from PMS to He flash (black line) and from ZAHB to thermal
pulses (green line). The figure is obtained as follows:
\begin{example}
> plot(track)
> plot(hbtk, add = TRUE, col = "green")
\end{example} 

\begin{figure}
\includegraphics[width=8cm]{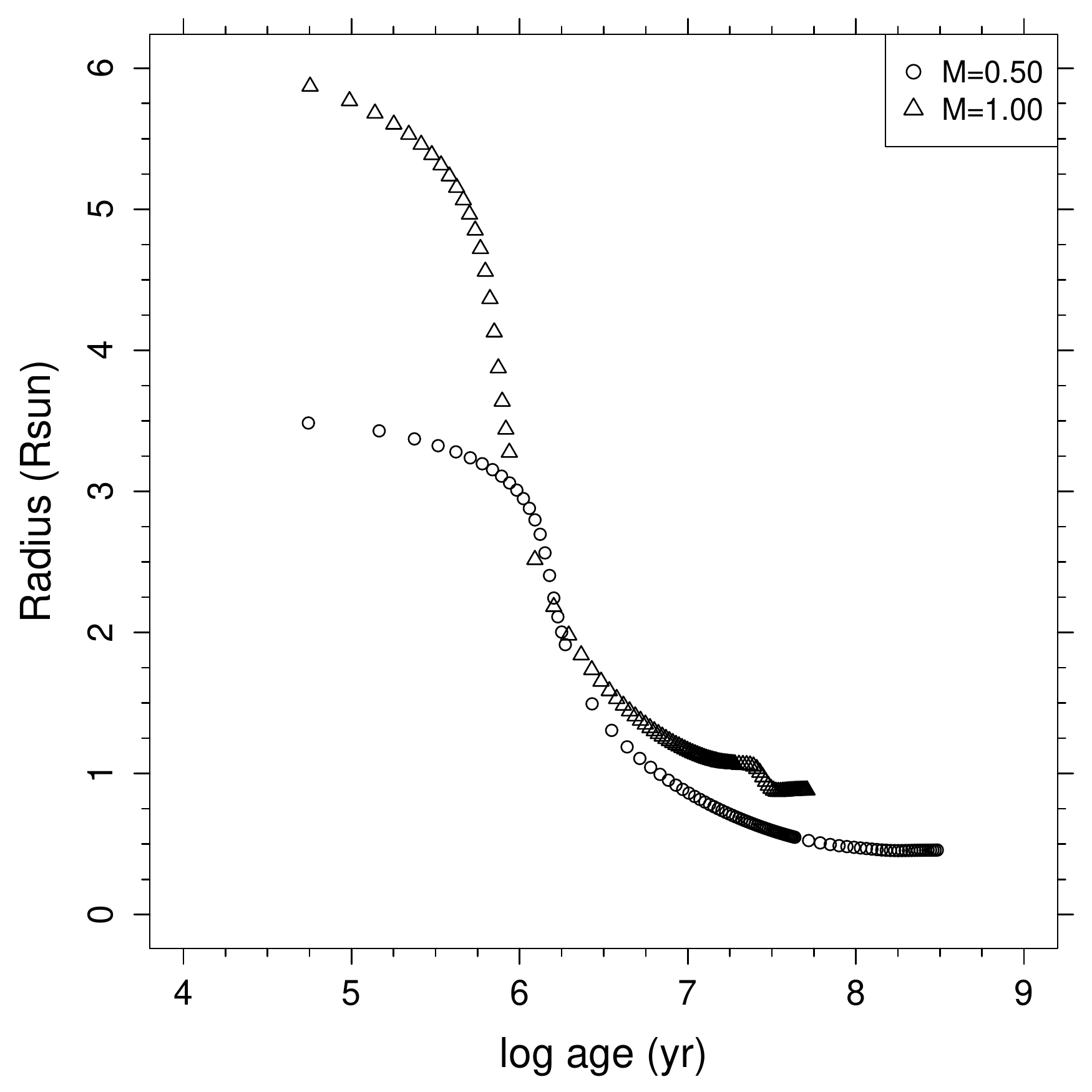}
\caption{\label{fig:radius}
Radius versus the logarithm of the age for the first 100 models
of two stars with different mass ($M$ = 0.50 $M_\odot$, and $M$ = 1.00
$M_\odot$) and identical composition, $z$ = 0.01, $y$ = 0.25, $\alpha_{\rm
ml}$ = 1.8, [$\alpha$/Fe] = 0.0.  }
\end{figure}

Apart from the plots discussed before, it is easy to display
other 
relation between the computed variables. In the following example we get the
data for two masses, namely 0.50 and 1.00 $M_\odot$ and plot the
trend of the radius (in units of solar radius) versus the logarithm of the age
for the first 100 models. The resulting plot is displayed in
Fig.~\ref{fig:radius}. 
\begin{example}
> trkr <- getTrkSet(m = c(0.5, 1), z = 0.01, 
                y = 0.25, ml = 1.8, afe = 0)
> mydata <- do.call(rbind, lapply(trkr, 
                "[[", "data"))
> D <- mydata[mydata\$mod <= 100, ]
> key <- as.numeric(factor(D\$mass))
> plotAstro(D\$time, D\$radius, type = "p", 
    pch = key, ylab = "Radius (Rsun)", 
    xlab = "log age (yr)")
> legend("topright", c("M=0.50", "M=1.00"), 
    pch = 1:2)
\end{example}

\begin{figure}
\includegraphics[width=8cm]{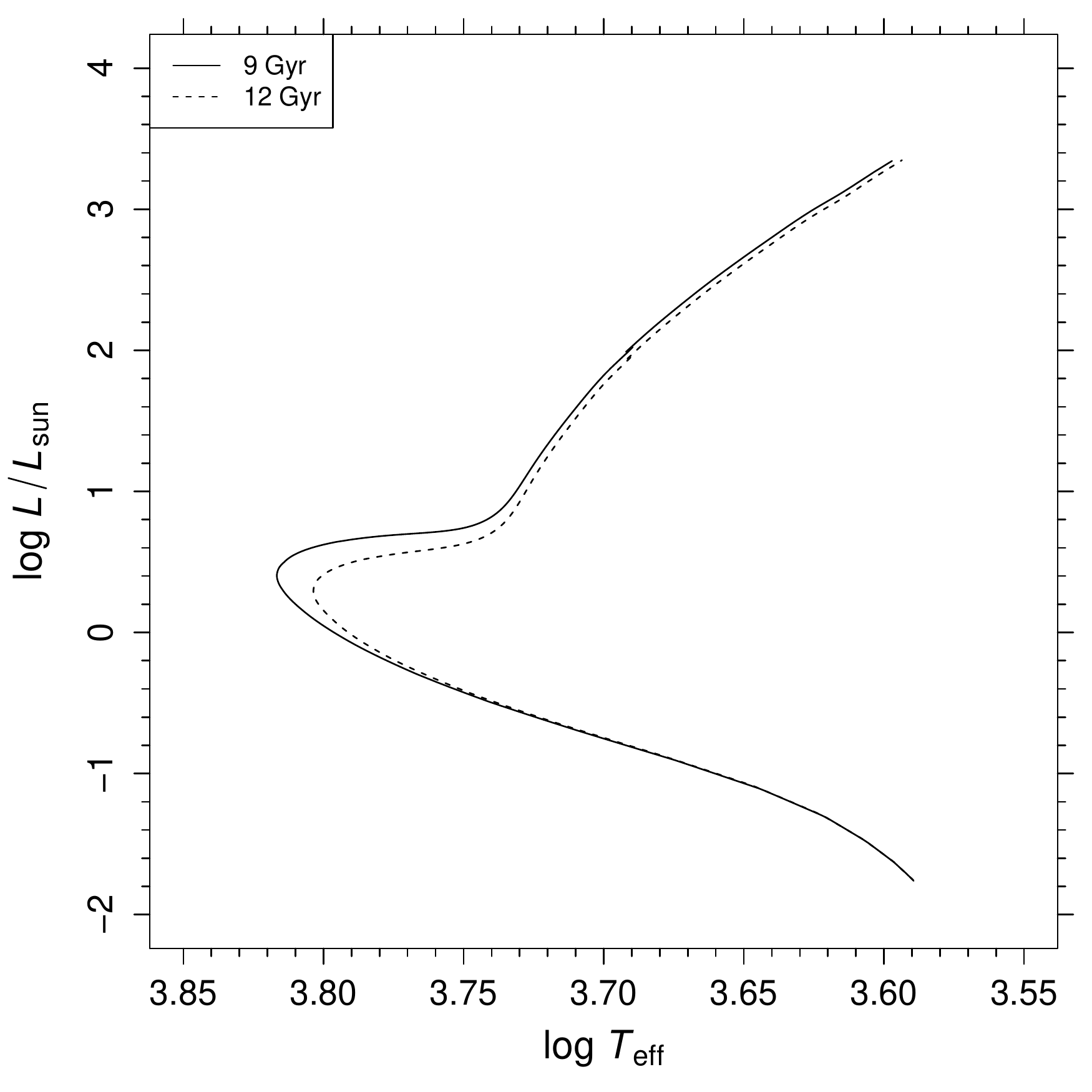}
\caption{\label{fig:iso}
Isochrones in the theoretical plane for 9 and 12 Gyr. Calculations performed
for $z$ =
0.001, $y$ = 0.25, $\alpha_{\rm ml}$ = 1.90, [$\alpha$/Fe] = 0.0. }
\end{figure}

Isochrones can be obtained from CDS and plotted in a similar way. As an
example, let get isochrones of 9 and 12 
Gyr for $z$ =
0.001, $y$ = 0.25, $\alpha_{\rm ml}$ = 1.90, [$\alpha$/Fe] = 0.0:
\begin{example}
> isc <- getIsoSet(age = c(9, 12), z = 0.001, 
             y = 0.25, ml = 1.90, afe = 0)
> isc
[[1]]
	 Stellar isochrone

Age = 9 Gyr
Z = 0.001 , Y = 0.25 
Mixing length = 1.9 
[alpha/Fe] = 0 

[[2]]
	 Stellar isochrone

Age = 12 Gyr
Z = 0.001 , Y = 0.25 
Mixing length = 1.9 
[alpha/Fe] = 0 

attr(,"class")
[1] "isoset"  "stellar"

> names(isc[[1]])
[1] "age"       "z"         "y"         "ml"        
    "alpha.enh" "data"

> names(isc[[1]]\$data)
[1] "logL" "logTe" "mass" "radius" "logg"  
\end{example}
The function returns an object of class \code{"isoset"}, a list containing
objects of class \code{"iso"}. These last objects are lists
containing metadata (age, metallicity, initial He abundance,
mixing-length, and $\alpha$-enhanchment) and the data frame \code{data}, which
contains the computed theoretical isochrones data. Figure \ref{fig:iso} shows
the set of isochrones plotted with the commands:
\begin{example}
> plot(isc, lty = 1:2)
> legend("topleft", c("9 Gyr", "12 Gyr"), 
         lty = 1:2)
\end{example}

\section{Tools for interpolating among data structures}

Even if the database provides a fine grid of models it is possible that the
specific model needed for a comparison with observational data has not been
calculated. To address this issue, the package \pkg{stellaR} includes tools 
for construction of new sets of isochrones.
The simplest case is whenever one desires an isochrone for ages not included in
the database, but for a combination of metallicity, He abundance,
mixing-length and $\alpha$-enhanchment existing in the database. 
The function \code{makeIso()} is able to compute the required
isochrones by mean of
interpolation on a tracks set. 
The interpolation can be performed for age in the range [7 - 15] Gyr. 
The user has the choice to explicitly provide to the function
a set of tracks, previously downloaded from CDS,  or to specify the required
composition of the tracks to be downloaded for the interpolation. 
To show the usage of the function we use the object \code{trks}
downloaded before to obtain an isochrone of age 9.7 Gyr: 
\begin{example}
> iso.ip <- makeIso(age = 9.7, tr = trks)
> iso.ip
	 Stellar isochrone

Age = 9.7 Gyr
Z = 0.001 , Y = 0.25 
Mixing length = 1.9 
[alpha/Fe] = 0 
\end{example}
the call produces a result identical to \code{makeIso(age = 9.7, z = 0.001, y
= 0.25, ml = 1.9, afe = 0)};  in this last
case the data are taken from CDS before the interpolation procedure.

The interpolation technique is based upon the fact that all
the tracks contain 
the same number of points by construction, and that a given point
corresponds to the same evolutionary phase on all the 
tracks. 
We define $S(M)$ as the set of tracks to be used for interpolation, parametrized
by the value of the mass $M$. Let $t_i(M)$ be the evolutionary
time for the $i$th point on the track of mass $M$, and $A$ be the age of the
required isochrone. Let $k$ be the point on the
track of lower mass of the set $S(M)$ for which $t_k(M) \geq A$.
For each point $j \geq k$ on the tracks in $S(M)$, a linear interpolation in
age of the values 
of mass, logarithm of the effective temperature and logarithm of the luminosity
is performed among tracks. These points define the required isochrone.
A potential problem of
this simple procedure will occur whenever massive
stars develop a convective core during the Main Sequence
(MS). In this case, as shown for example in
\citet{ginevra2012}, the monotonic trend of the evolutionary time --
that decreases with increasing stellar mass at
the end of the MS -- inverts at the middle
of the MS. However the problem will be encountered
only for early-time isochrones, for which the mass at the
isochrone Turn-Off will be in the interval during which the convective
core develops. The procedure outlined
in this Section is adequate for construction of
isochrones throughout the range allowed by the function.

\subsection{Tracks interpolation}

The package \pkg{stellaR} provides also a tool for performing a 3D
interpolation 
on the database to construct a set of tracks for values of metallicity,
helium abundance and mixing-length not included in the computations available
at CDS. 
The function \code{interpTrk()} can be used for this procedure.
A call to this function causes the download from CDS of the sets of tracks
needed for the interpolation. 

The new set of tracks is computed by mean of a linear
interpolation. The metallicity
is log-transformed before the interpolation procedure.
Let $T_{z,y,\alpha_{\rm ml}}(M)_i$ be $i$th point on the dataset containing the
evolutionary time, the effective temperature and the logarithm of surface
luminosity for the track of mass $M$, and given composition. 
The interpolation algorithm proceeds as follows: 
\[ \overbrace{T_{z,y,\alpha_{\rm ml}}\left(M\right)_i}^{\rm 8 \; sets} \rightarrow 
\overbrace{T_{z,y,*}\left(M\right)_i}^{\rm 4 \; sets} \rightarrow
\overbrace{T_{z,*,*}\left(M\right)_i}^{\rm 2 \; sets} \rightarrow 
\overbrace{T_{*,*,*}\left(M\right)_i}^{\rm 1 \; sets} \]
where the symbol $*$ means that interpolation occurred in the substituted
variable.  
The selection of the tracks set which enter in the interpolation is based upon 
the identification of the vertexes of the cell of the ($z$, $y$, $\alpha_{\rm
ml}$) space containing the point identified by the required parameters.
Then, for all the 17 masses at the vertexes, the linear interpolation
described above is performed.   
In the worst case scenario, whenever no one of the supplied parameters values
exist in the database, the 
interpolation requires $2^3 = 8$ sets of 17 masses.
The algorithm is however able to reduce the dimensionality of the process if
some of 
the variables values exist in the database. 

As a demonstration, let us compute a set of tracks with mixing-length value
$\alpha_{\rm ml}$ = 1.74, $z$ = 0.002, $y$ = 0.25, [$\alpha$/Fe] = 0.0:
\begin{example}
> ip.trk <- interpTrk(z = 0.002, y = 0.25, 
                       ml = 1.74, afe = 0)
\end{example}
Since the values of $z$ and $y$ exist in the database, only an interpolation
on the mixing-length value is performed by the code. 
The set of tracks can be used for isochrones construction, like a standard set
of tracks:
\begin{example}
> ip.iso <- makeIso(age = 12, tr = ip.trk)
\end{example}

\section{Keypoints extraction}

Important stages in the evolution of a star are defined as "keypoints",
e.g.\ hydrogen core exhuastion, Turn-Off luminosity, RGB tip luminosity.
To simplify their extraction the
package \pkg{stellaR} provides the function \code{keypoints()}, which
operates 
on an object of class \code{"trk"} or \code{"iso"}.

The function extracts from the data stored in objects of class \code{"trk"}
the 
rows of the data frame relative to the following evolutionary stages:
\begin{enumerate}
\item{ZAMS. Zero-Age Main-Sequence, defined as the point for which the
  central H abundance drops below 99\% of its initial value. }
\item{TO. Turn-Off, defined as the point for which the effective
  temperature reaches its maximum value. If multiple lines satisfy the
  constraint, the values of all the rows are averaged. }
\item{BTO. Brighter Turn-Off, defined as the point for which the
  effective temperature drops below the one of the TO minus 100 K. The
  points can not exist for low masses. Details on the  advantages of this
  evolutionary point with respect to the TO can be found
  in \citet{Chaboyer1996}.} 
\item{exHc. Central H exhaustion, defined as the point for which the central
  H abundance is zero. For low masses the point can coincide with
  TO. This is the last point of the tracks with mass lower or equal to
  0.50 $M_\odot$.}
\item{Heflash. Helium flash, the last point of the track for masses
  higher than 0.50  $M_\odot$.}
\end{enumerate}
When  the function is called on an object of class \code{"iso"} it returns a
data frame containing only TO and BTO phases.

In both cases the function inserts in the returned data frame the columns
relative to mass (or age for isochrones), 
metallicity, initial He value, mixing-length, $\alpha$-enhancement, and
evolutionary phase identifier.

As a demonstration we extract the TO and BTO points from the
object \code{isoc} generated in a previous example:
\begin{example}
> kp <- keypoints(isoc)
> kp
          logL    logTe      mass  radius 
TO   0.4044723 3.816652 0.8396903 1.23558
BTO  0.5556971 3.809943 0.8561162 1.51671
TO1  0.2917250 3.803611 0.7772973 1.15234
BTO1 0.4495597 3.796545 0.7909500 1.42768
         logg age     z    y  ml alpha.enh 
TO   4.178911   9 0.001 0.25 1.9         0
BTO  4.009265   9 0.001 0.25 1.9         0
TO1  4.205965  12 0.001 0.25 1.9         0
BTO1 4.027426  12 0.001 0.25 1.9         0
     id
TO    1
BTO   2
TO1   1
BTO1  2
\end{example}

\begin{figure}
\includegraphics[width=8cm]{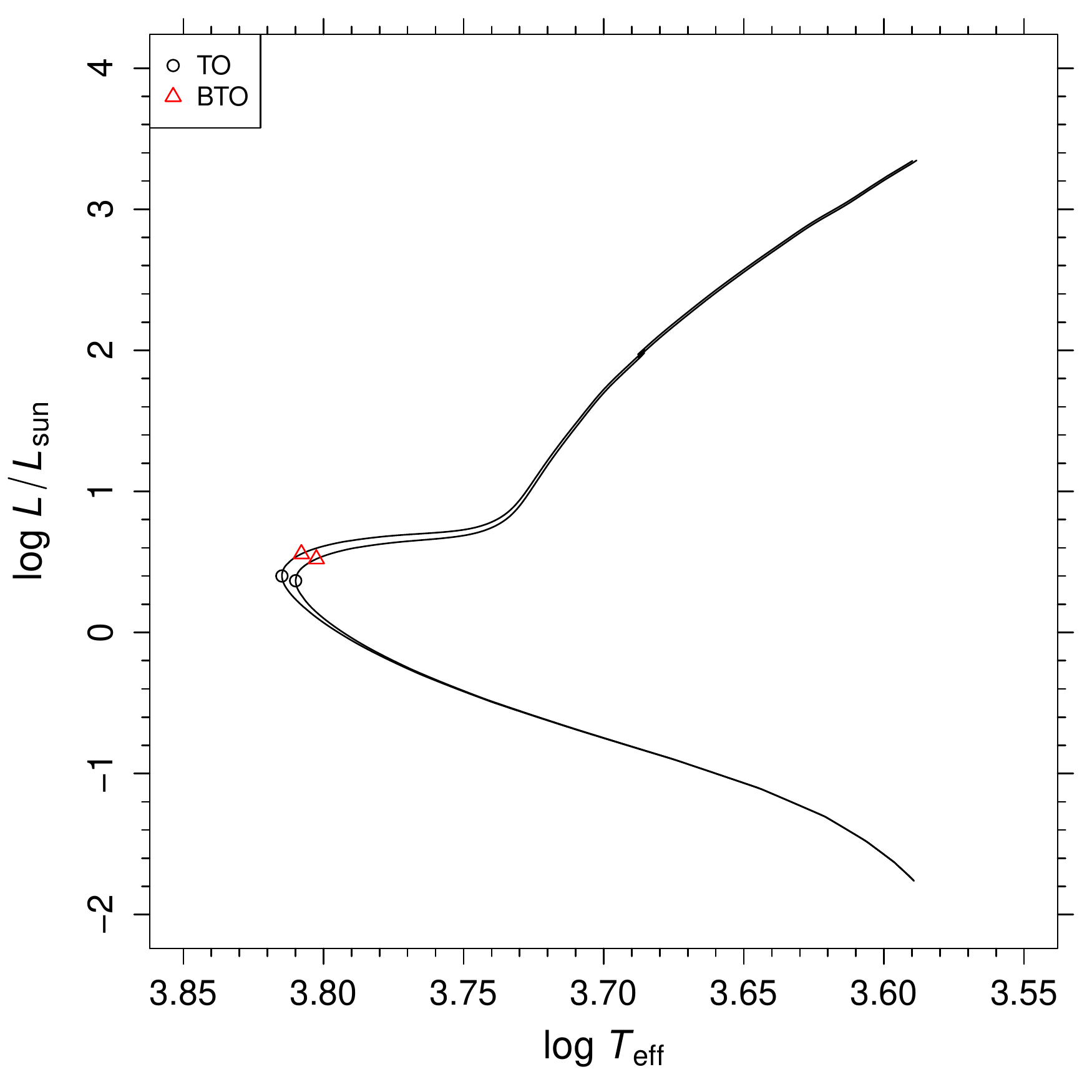}\\
\includegraphics[width=8cm]{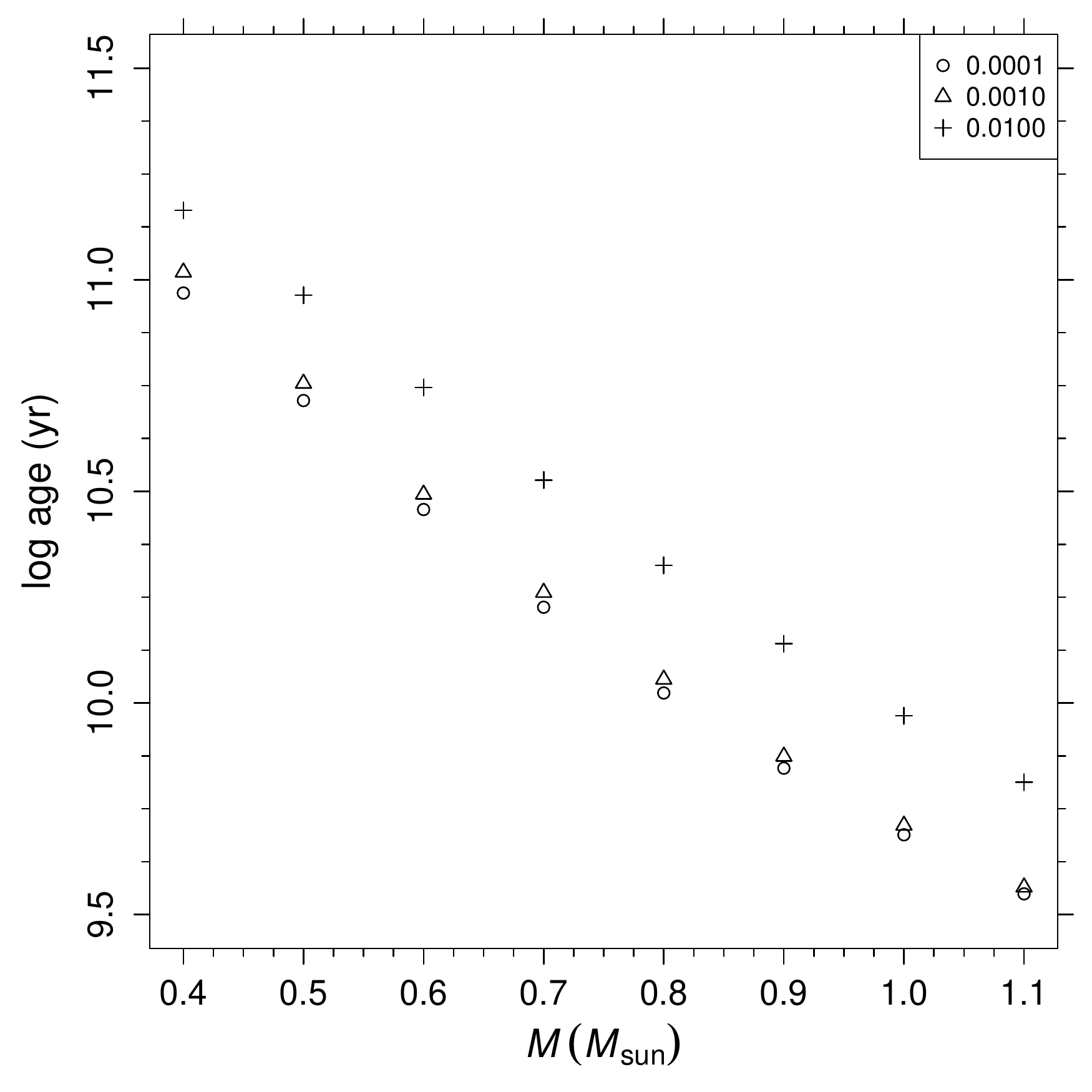}
\caption{\label{fig:keypoints}
Top panel: same as Figure \ref{fig:iso}. The position of Turn-Off (TO) and
Brighter  TO (BTO) are shown. Bottom panel: time to exhaustion of central
hydrogen as a function of the mass of the star. The symbols identify the
values of the metallicity $z$ form 0.0001 to 0.01.}
\end{figure}

The points can be easily superimposed to the isochrones in the theoretical
plane. The top panel of Figure \ref{fig:keypoints} is obtained with the
following commands: 
\begin{example}
> plot(isoc)
> points(logL ~ logTe, data = kp, pch = id, 
               col = id)
> legend("topleft", c("TO", "BTO"), 
         pch = 1:2, col = 1:2)
\end{example}
As a last example we extract a set of tracks for masses  in the range [0.40 -
1.10] 
$M_\odot$ and three metallicity $z$ = 0.0001, 0.001, 0.01 and we display
(bottom panel in Figure \ref{fig:keypoints}) the time
of exhaustion of central hydrogen as a function of the mass of the star. 
\begin{example}
> mT <- seq(0.4, 1.1, by = 0.1)
> zT <- c(0.0001, 0.001, 0.01) 
> tr <- getTrkSet(m = mT, z = zT, y = 0.25, 
        ml = 1.9, afe = 1)
> kp <- keypoints(tr)
> kpH <- kp[kp\$id == 4,]

> xlab <- expression(italic(M) ~ 
          (italic(M)[sun]))
> ylab <- "log age (yr)"
> symbol <- as.numeric(factor(kpH\$z))
> plotAstro(kpH\$M, kpH\$time, type = "p", 
   pch = symbol, xi = 0.1, xlab = xlab, 
   ylab = ylab)
lab <- levels(factor(format(kpH\$z, 
         nsmall = 4, scientific = FALSE)))
> legend("topright", lab, pch = 1:3)
\end{example}

\section{Summary}

This paper demonstrated how the package \pkg{stellaR} can be useful to the
astrophysical community as a tool to simplify the access to stellar tracks and
isochrones calculations available on-line. A set of tools to access, manipulate
and plot data are 
included in the package and their usage was shown. The interpolation
functions included in the package can be used to safely produce tracks or
isochrones at compositions not included in the database, without the need that
users develop software on their own.
A planned extension of the package is the modification of the algorithm of
isochrone construction to make feasible the calculation of isochrone of young
ages. This step can be useful in view of a possible extension of the Pisa
database to higher masses, or to manipulation of data stored in other
databases. In fact, while   
the package is currently developed for accessing data from Pisa
low-mass database, other public databases can be in principle accessed
in the same way. 
This step is however complicated by the fact that no standard for stellar
model output exists in the astrophysical community, requiring the
personalization  of
the interface functions for each set of data.

\section{Acknowledgments}
We are grateful to our anonymous referees for many stimulating suggestions that
helped in clarify and improve the paper and the package. We thank Steve
Shore for a careful reading of the manuscript and many useful suggestions. 

\bibliography{biblio}

\address{Matteo Dell'Omodarme\\
  Dipartimento di Fisica ``Enrico Fermi'',
  Universit\`a di Pisa\\
  Largo Pontecorvo 3, Pisa I-56127 \\
  Italy}\\
\email{mattdell@fastmail.fm}

\address{Giada Valle\\
  Dipartimento di Fisica ``Enrico Fermi'',
  Universit\`a di Pisa\\
  Largo Pontecorvo 3, Pisa I-56127 \\
  Italy}\\
\email{valle@df.unipi.it}

\end{article}

\end{document}